\documentclass[aps,pra,twocolumn,floats]{revtex4}

\usepackage{dcolumn}
\usepackage{graphicx}
\usepackage{epsfig}

\newcommand {\ket}[1]{|{#1}\rangle}

\begin{document}

\title{Non-Markovian spontaneous emission dynamics of a quantum emitter near a MoS$_2$ nanodisk}

\author{Ioannis Thanopulos$^{1}$}


\author{Vasilios Karanikolas$^2$}


\author{Nikos Iliopoulos$^2$}


\author{Emmanuel Paspalakis$^{2}$}


\affiliation{$^1$ Department of Optics and Optometry, T.E.I. of Western Greece, Aigio 251 00, Greece}

\affiliation{$^2$ Materials Science Department, School of Natural Sciences,
University of Patras, Patras 265 04, Greece}

\begin{abstract}
We introduce a photonic nanostructure made of two dimensional materials that can lead to non-Markovian dynamics in the spontaneous emission of a nearby quantum emitter. Specifically, we investigate the spontaneous emission dynamics of a two-level quantum emitter with picosecond free-space decay time, modelling J-aggregates, close to a  MoS$_2$ nanodisk. Reversible population dynamics in the quantum emitter is obtained when the emitter's frequency matches the frequency of an exciton-polariton resonance created by the nanodisk. When such isolated resonances exist, decaying Rabi oscillations may occur. The overlapping of exciton-polariton resonances also affects strongly the decay dynamics at close distances to the nanodisk, giving rise to complex decaying population oscillations. At very close distances of the emitter to the nanodisk the ultrastrong coupling regime appears, where after a very fast oscillatory partial decay of the initial population, the emitter rest population remains constant over long times and population trapping occurs. The size and material quality of the nanodisk is shown to be of lesser influence on the above results.
\end{abstract}

%

\maketitle


\section{Introduction}

Strong coupling between a quantum emitter (QE) and its photonic environment is a distinct regime of light-matter interaction, which manifests itself in coherent oscillations of energy between matter and the photonic subsystem leading to non-Markovian effects in spontaneous emission. Besides fundamental interest, strong coupling also holds a great potential for various useful applications in quantum technologies, all-optical logic, and control of chemical reactions \cite{baranov18}.

The prototype physical system for obtaining non-Markovian effects in spontaneous decay consists of a two-level QE coupled to a photonic reservoir with a strongly modified density of modes \cite{ReviewnonMarkov,Lodahl15a}. Examples of photonic structures that give non-exponential spontaneous decay include microwave and optical cavities \cite{cqed},  photonic crystals \cite{pbg1,pbg2}, semiconductor microcavities \cite{Lodahl11a}, carbon nanotubes \cite{Bondarev04a}, two-dimensional structured reservoirs \cite{Tudela17a}, zero-index structures \cite{Liberal17a}, and plasmonic nanostructures \cite{Tejedor14,Hakami14,Varguet16a,Li16a,thano17,Yang17a,iliopoulos18,Neuman18a,Rousseaux18a}.

Transition metal dichalcogenide (TMD) monolayers \cite{Kormanyos2015} are atomically thin, direct bandgap  semiconducting two-dimensional materials featuring bandgaps in the visible
and near-infrared range, strong excitonic resonances, high
oscillator strengths, and valley-selective response; they support
exciton polaritons, too \cite{Basov16a}. TMDs are usually used as emission sources or absorbing layers \cite{Messer2015}
in combination with noble metallic layers
\cite{Kang2017}, metallic nanodisks \cite{Dutta2018}, and photonic
crystals \cite{Ge2018}. Additionally, the optical properties of QEs,
such as molecules or quantum dots, near single or multilayer TMDs, or their nanostructures, have been investigated.
In such cases the spontaneous decay of the QE is modified due to the Purcell effect and enhanced energy transfer, as well as, ultrafast charge transfer may occur \cite{Prins2014,Chen,Zang2016,Goodman18a,Karanikolas2016a,paspalakis17}.

The studies of QEs next to TMDs remain in the weak coupling regime, where a Markovian response of the QE holds and exponential spontaneous decay occurs. Non-Markovian dynamics of spontaneous emission next to a TMD monolayer or a TMD nanostructure remains unexplored. Non-Markovian effects in spontaneous emission arise when the density of modes of the photonic structure presents strong and narrow resonances. Confining the transverse magnetic exciton-polariton mode from a MoS$_2$ layer to a MoS$_2$ nanodisk leads to the redistribution of the available modes and the creation of localized exciton-polariton modes. We have shown that there is significant enhancement of the Purcell factor of the QE combined with ultranarrow resonances \cite{paspalakis17}, which shows that MoS$_2$ nanodisks are viable candidates for creating reversible spontaneous emission dynamics in QEs. The purpose of this Letter is to explore this possibility.

Here, we combine quantum dynamics calculations invoking and beyond the rotating wave approximation (RWA) with electromagnetic calculations and show that the decay dynamics of a two-level QE, modelling J-aggregates,  close to a MoS$_2$ nanodisk  can exhibit strong non-Markovian effects. We consider QEs located at different distances from a 30 nm and 7.5 nm radius MoS$_2$ nanodisk with resonant frequencies close to the excitonic resonances of the MoS$_2$. Depending on the actual resonant frequency, the distance from the MoS$_2$ nanodisk and the free-space decay rate of the QE, the excited state population may exhibit decaying Rabi oscillations, complex decaying oscillations, and even ultrastrong coupling and population trapping. We also investigate the influence of the material quality of the MoS$_2$ disk on
the quantum dynamics by considering high-, as well as low-quality MoS$_2$ material.

\section{Theory}
\begin{figure}[h]
\centerline{\hbox{\epsfxsize=70mm\epsffile{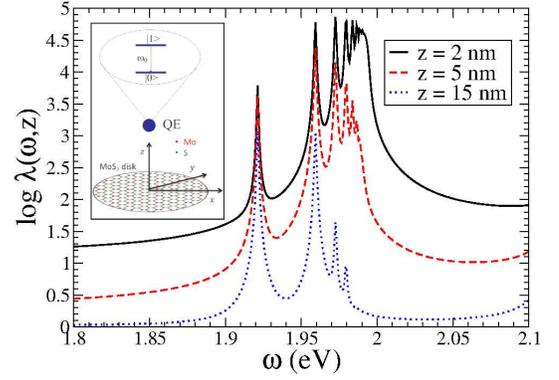}}}
\begin{center}
\caption{\label{fig1} (color online) Purcell factor $\lambda(\omega,z)$ of a QE with $x$-oriented transition dipole moment located at $\mathbf{\vec{r}}_{\text{QE}}=(0,0,z)$ nm next to a high-quality MoS$_2$ nanodisk of radius $R=30$ nm.  Inset: The two-level QE with free-space resonance frequency $\omega_0$ placed at a distance $z$ perpendicular to the center of a nanodisk of radius $R$.}
\end{center}
\end{figure}

We investigate the non-Markovian dynamics of the spontaneous emission of a single photon QE interacting with a MoS$_{2}$ nanodisk. We consider a two-level QE located at
${\bf \vec{r}}=(0,0,z)$ of a coordinate system with origin at the center of the nanodisk, as shown in the inset in
Fig. \ref{fig1}. The state of the system is given by
\begin{eqnarray}\label{psistate}
\ket{\Psi(t)}&=&c_1(t) e^{-i\omega_0^\prime t}\ket{1;0_\omega} + \nonumber \\
&&\int d\vec{r}\int d\omega~ c(\vec{r},\omega,t) e^{-i\omega t}\ket{0;1_{\vec{r},\omega}} \, .
\end{eqnarray}
Here, $\ket{n;a} = |n\rangle \otimes |a\rangle$, where $|n\rangle$ $(n=0, 1)$ denotes the quantum states of
the two-level system, see Fig. \ref{fig1}, and $|a\rangle$ denotes the photonic states of the modified electromagnetic  modes due to the presence of the nanodisk, with $\ket{0_\omega}$ standing for the vacuum
and $\ket{1_{\vec{r},\omega}}$ for one photon states.
The equation for $c_1(t)$ is given by  (we use $\hbar = 1$ in this paper)
\begin{eqnarray}\label{eq0}
\dot{c}_1(t) & = & i\int_0^t  K(t-t^\prime) c_1(t^\prime) dt^\prime  \, , \\
K(t-t^\prime) & = & ie^{i\omega_0^\prime(t-t^\prime)}\int_0^\infty \tilde{J}(\omega)
e^{-i\omega(t-t^\prime)} d\omega \label{kernel} \, ,
\end{eqnarray}
with $\tilde{J}(\omega) = J(\omega)\bigg(\frac{\omega}{\omega_0}\bigg)^3$, where $J(\omega)=(\frac{2\omega_0}{\omega_0+\omega})^2 \Gamma_0(\omega_0)\lambda(\omega,z)/2\pi$
and $J(\omega)=\Gamma_0(\omega_0)\lambda(\omega,z)/2\pi$ with and without taking into account
counter-rotating effects, respectively \cite{iliopoulos18,thano17}.
In Eqs. (\ref{psistate})-(\ref{kernel}),  we use $\omega_0^\prime=
\omega_0-\Delta\omega_{ndyn}$, with $\Delta\omega_{ndyn}=(1/2\pi)^{-1}\int_0^\infty \frac{2\omega^3}{\omega_0^2(\omega_0+\omega)^2}\Gamma_0(\omega_0)\lambda(\omega,z)d\omega$ standing for the relative energy shift between the two levels in case counter-rotating effects are considered \cite{iliopoulos18}, or
$\Delta\omega_{ndyn}=0$ otherwise;  $\Gamma_0(\omega_0)$ is the free-space decay width of the QE with free-space resonance frequency $\omega_0$, and $\lambda(\omega,z)$ is the Purcell factor due to the location of the QE at $\mathbf{\vec{r}}_{\text{QE}}=(0,0,z)$ close to the nanodisk.  We compute the probability amplitude dynamics of the QE $c_1(t)$ by applying the effective mode differential equation methodology \cite{thano17}.

In order to quantify the influence of the modified environment on the QE emission we calculate the
Purcell factor, which is defined as:
\begin{equation}
\lambda(\omega,\mathbf{\vec{r}})=
\sqrt{\varepsilon}+\frac{6\pi c}{\omega}\hat{n}\cdot\mathrm{Im}\,\hat{\bf G}(\mathbf{\vec{r}},\mathbf{\vec{r}},\omega)\cdot\hat{n} , \label{eq:01}
\end{equation}
where $\varepsilon$ is the permittivity of the host medium,
$\hat{\bf G}$ is the induced part of the electromagnetic  Green's tensor, due to the nanodisk, calculated at the QE position, which represents the response of the nanodisk geometry under consideration to a point-like dipole excitation \cite{Dung2000}, and $\hat{n}$ is the unit vector along the direction of the transition dipole moment. We deal with the near field regime of the QE, for which the QE separation from the MoS$_{2}$ nanodisk is much smaller than the spontaneous emission  wavelength of the QE, $\left|\mathbf{r}\right|\ll\lambda$.
Thus, we can use the electrostatic approximation.

In this work we calculate the Purcell factor of the QE due to the presence of the nanodisk by application of Green's tensor methods \cite{Karanikolas2016a,paspalakis17}. We consider a free-standing MoS$_{2}$ disk, with host medium dielectric permittivity $\varepsilon=1$. The transition dipole moment of the
QE is along the $x$ direction, i.e. parallel to the nanodisk,  and the QE is always located on the axis of rotational symmetry of the nanodisk, implying $\mathbf{\vec{r}}_{\text{QE}}=(0,0,z)$. The induced Green's tensor is given by the expression \cite{Karanikolas2016a,paspalakis17}:
\begin{equation}
\hat{\bf G}_{xx}(z,\omega)=-\frac{c^{2}}{2\omega^{2}}
\sum_{n=0}^{\infty}c_{n}^{1}(z,\omega)\frac{\bigg[\tilde{\cal R}-z/R\bigg]^{2n+2}}
{\tilde{\cal R}},\label{eq:02}
\end{equation}
with $\tilde{\cal R}=\sqrt{(z/R)^{2}+1}$.
The expansion coefficients $c_{n}^{1}(z,\omega)$  are obtained as solutions
of a matrix equation, where their values depend on the position of
the QE, the angular eigenmode $l=1$ and the radial eigenmode $n$ \cite{paspalakis17}.

MoS\textsubscript{2} is a direct gap semiconductor with relatively
intense photoluminescence. The resonance part of the two-dimensional
optical conductivity of the MoS\textsubscript{2}, $\sigma_{\text{res}}$,
takes into account the interaction of light with the lowest energy
$A$ and $B$ excitons \cite{paspalakis17} and is given by
\begin{equation}
\sigma_{\text{res}}(\omega)=\frac{4\alpha_{0} cv^{2}}{\pi a_{\text{ex}}^{2}\omega}
\sum_{k=A,B}\frac{-i}{\omega_{k}-\omega-i\gamma_{k}},\label{eq:07-1}
\end{equation}
where $\alpha_{0}$ is the fine structure constant, $a_{\text{ex}}=0.8\,\text{nm}$
is the exciton Bohr radius, the damping parameters are $\gamma_{A}$
and $\gamma_{B}$, and the exciton energies are $\omega_{A}=1.9\,\text{eV}$
and $\omega_{B}=2.1\,\text{eV}$. $v$ is a constant velocity, which is
connected with the hopping parameter, and for MoS$_2$
we use the value $v=0.55$ {nm/fs} \cite{Xiao2012,Zhang2014d}.
This optical conductivity includes only the contribution from the
bright direct excitons, which dominate due to their large oscillator
strength.
We note that $\gamma_{\text{A}}$ and $\gamma_{\text{B}}$,
are related to the quality of the MoS\textsubscript{2} material
which depends on different temperatures and different preparation methods.
In particular, here we consider high-quality material with  $(\gamma_{\text{A}}=0.5~\mbox{meV},\gamma_{\text{B}}=1.1~\mbox{meV})$ and low-quality material with
$(\gamma_{\text{A}}=2.5~\mbox{meV},\gamma_{\text{B}}=5.6~\mbox{meV})$.

At higher energies, the interband transitions need to be included
in the model describing the surface conductivity. We model these transitions
with an expression of the form
\begin{equation}
\text{\text{Re}}\left[\sigma_{\text{inter}}\right]=\sigma_{0}
\tilde{\cal E}
\left[1+\frac{1+2\omega_{B}\beta}{\Omega^{2}}\left(1+\omega_{B}\beta-{\cal E}
\right)\right],\label{eq:08-1}
\end{equation}
where $\tilde{\cal E}=m\Theta(\omega-\omega_B)/{\cal E}$, ${\cal E}=\sqrt{1+2\omega_B\beta+\Omega^2}$, $\Theta(\cdot)$ is the Heaviside step function,
$\Omega=\omega/\omega_{B}$
and $\beta$ is a mixing parameter; for MoS$_2$ $\omega_{B}\beta=0.84$ \cite{Stauber2015a}.
The parameter $m$ is for absorption scaling; here, the value $m=1$ is used.

In Fig. \ref{fig1}  we explicitly show the Purcell factors $\lambda(\omega,z)$ along the $x$ direction  for a QE located at $z = 2, 5, 15$ nm from a high-quality MoS$_2$ nanodisk of radius  $R=30$ nm. A significant enhancement of the Purcell factor is observed for over four orders of magnitudes. The strong and sharp peaks represent the exciton-polariton resonances. In this work, we study the decay dynamics for  state-of-the-art two-level QEs with transition frequencies in the optical regime, such as J-aggregates with free-space decay time in the ps time regime \cite{baranov18,knoester90}. We note that the J-aggregate size, which is on the order of
a few nanometers, renders the dipole approximation used in our Hamiltonian a good approximation for photon energies around 1.95 eV (corresponding wavelength $\approx$ 635 nm).

\section{Results}
\begin{figure}[h]
\centerline{\hbox{\epsfxsize=70mm\epsffile{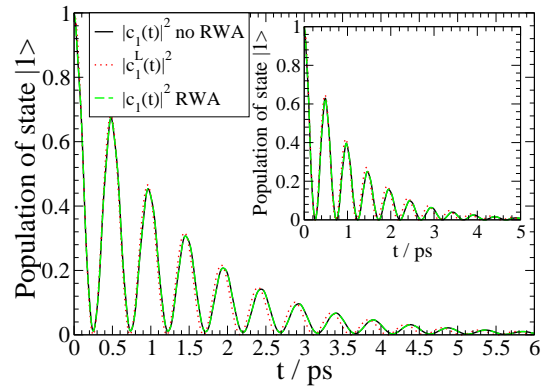}}}
\begin{center}
\caption{\label{fig2} (color online) Population dynamics of a two-level QE with a $x$-oriented transition dipole moment and $\Gamma_0(\omega_0)=59$ $\mu$eV; the QE is located at $\mathbf{\vec{r}}_{\text{QE}}=(0,0,15)$ nm next to a high-quality MoS$_2$ disk with radius $R=30$ nm. A QE with $\omega_0=1.92128$ eV (main panel) and
$\omega_0=1.95945$ eV (inset) are presented.  In each panel, the $|c^L_1(t)|^2$ (red dotted curve) and the
$|c_1(t)|^2$ without the RWA (black solid curve)  and with the RWA (green dashed curve) are shown; we note that the curves corresponding to dynamics with and without the RWA are practically indistinguishable.
Each $|c^L_1(t)|^2$ curve is obtained by assuming $\omega_c=\omega_0$.  See text for discussion.}
\end{center}
\end{figure}

In Fig. \ref{fig2} we present the population dynamics $|c_1(t)|^2$  without
the RWA (black solid curve) and with the RWA (green dashed curve) for a QE located at $z=15$ nm from a high-quality MoS$_2$ nanodisk with $\Gamma_0(\omega_0)=59$ $\mu$eV. We investigate a QE with $\omega_0=1.92128$ eV (main panel) and $\omega_0=1.95945$ eV (inset); these frequencies correspond to the two strongest exciton-polariton resonances of the nanodisk at this distance.  In both cases, the $|c_1(t)|^2$ with and without invoking the RWA, which are practically indistinguishable in this case, shows distinct non-Markovian features. We find that the population of the upper level of the QE has Rabi oscillations with gradually decreasing amplitude before it totally decays out of the QE within 6 ps. We note that at this distance, although the coupling of the QE to the electromagnetic  mode continuum is strong enough in order to observe clear non-Markovian features in the spontaneous emission  dynamics, the influence of counter-rotating effects is marginal, since the  $|c_1(t)|^2$ curves in both cases are indistinguishable. We also note that the influence of the counter-rotating effects on the spontaneous emission  dynamics is directly related to the value of $\Delta\omega_{ndyn}$, which is less than 10 $\mu$eV for both frequencies at this distance.

In the case that $J(\omega)$ is approximated by a Lorentzian form \cite{Petruccione},
\begin{equation}
J^L(\omega) = \frac{1}{2\pi}\frac{\Gamma_0(\omega_{0})\lambda(\omega_0,z)\beta^2}{(\omega_{0}-\omega-\Delta)^2 + \beta^2},
\end{equation}
with $\Delta\equiv\omega_0-\omega_c$ being the detuning between the QE frequency $\omega_{0}$ and a cavity-like central mode frequency $\omega_c$, and $\beta$ denoting the spectral width of the coupling to the central mode, by taking the limit of the integral in Eq. (\ref{kernel}) to $-\infty$ and using the Laplace
transform, the upper level population dynamics of the QE  can be now calculated analytically as
\begin{equation}\label{dynam}
c^L_{1}(t) = e^{-0.5\tilde{\beta}t}\left[\cosh\left(\frac{q t}{2}\right)+\frac{\tilde{\beta}}{q}\sinh\left(\frac{q t}{2}\right)\right] \,,
\end{equation}
with $q = \sqrt{\tilde{\beta}^2 - 2\Gamma_0(\omega_{0})\lambda(\omega_0,z)\beta}$ and $\tilde{\beta}=\beta-i\Delta$. Note that Eq. (\ref{dynam}) is derived
within the flat-continuum approximation \cite{thano17} and the RWA \cite{Petruccione}.

In Fig. \ref{fig2}, for each $\omega_0$, we also present the upper level population dynamics of the QE
$|c^L_1(t)|^2$ (red dotted curve) after fitting the $\tilde{J}(\omega)$ for a QE at $z=15$ nm using the  Lorentzian profile $J^L(\omega)$ with $\omega_c=\omega_0$ \cite{betas}. In both panels, the population dynamics $|c^L_1(t)|^2$  is almost indistinguishable from the dynamics computed by using $\tilde{J}(\omega)$
as obtained from the electromagnetic Green's tensor calculations; a clear indication that at this distance and frequency, the nanodisk affects the
spontaneous emission  dynamics as a single-mode cavity with central frequency at the exciton-polariton resonance of the nanodisk.

\begin{figure}[h]
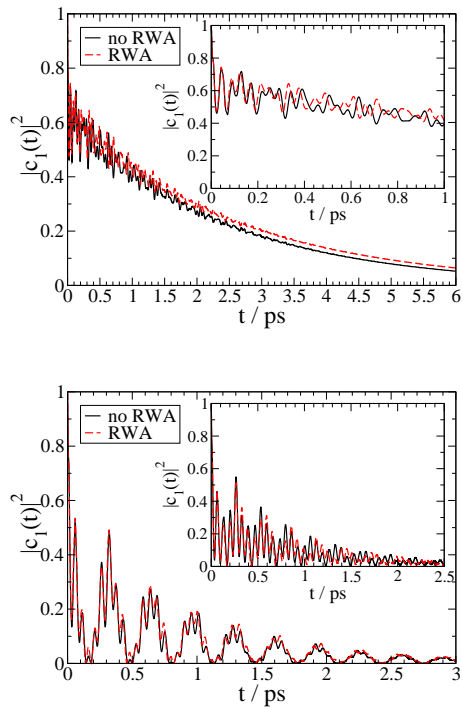

\centerline{\hbox{\epsfxsize=60mm\epsffile{fig3a.eps}}}
\vspace*{0.7cm}
\centerline{\hbox{\epsfxsize=60mm\epsffile{fig3b.eps}}}
\begin{center}
\caption{\label{fig3} (color online)   Population dynamics  with and without the RWA of a QE with a $x$-oriented transition dipole moment and $\Gamma_0(\omega_0)=59$ $\mu$eV located at $\mathbf{\vec{r}}_{\text{QE}}=(0,0,5)$ nm next to a high-quality MoS$_2$ disk  with radius $R=30$ nm. Upper panel: A QE with   $\omega_0=1.92128$ eV. The inset presents the dynamics as in the main panel at early times. Lower panel: A QE with
$\omega_0=1.959456$ eV (main panel) and $\omega_0=1.97262$ eV (inset).}
\end{center}
\end{figure}

The dynamics with and without invoking the RWA for a QE with $\Gamma_0(\omega_0)=59$ $\mu$eV located at $\mathbf{\vec{r}}_{\text{QE}}=(0,0,5)$ nm next to a high-quality MoS$_2$ nanodisk is presented in Fig. \ref{fig3}. In the upper panel, we show the $|c_1(t)|^2$ time evolution for a QE with $\omega_0=1.92128$ eV; the inset in this panel presents the same dynamics  at early times. Independently whether the RWA is invoked or not, we observe population decay, which is practically completed within 7 ps, with small-amplitude, high-frequency, oscillatory features superimposed on it. These high-frequency oscillatory features are due to the fact that at this distance the exciton-polariton resonance at $\omega=1.92128$ eV is overlapping significantly with the next one at
$\omega=1.959456$ eV. However, although the frequency of the QE corresponds nominally to an exciton-polariton resonance here, the observed dynamics resembles rather to a QE with an off-resonance frequency with respect to the exciton-polariton resonance at $\omega=1.92128$ eV. This can be understood as a result of the dynamical shift induced on the QE transition frequency by the presence of the modified by the nanodisk electromagnetic mode continuum; on top of it, we also have the shift due to the counter-rotating effects in this case, which  here
amounts to $\Delta\omega_{ndyn}\approx 0.3$ meV. We also observe that the population dynamics without the RWA is quantitatively different from the dynamics when the RWA is invoked, while the overall qualitative dynamical features remain the same in both cases.  These observations lead us to the conclusion that the
off-resonant behaviour of the QE dynamics here are marginally depending on invoking the RWA or not. However, the observation that after a short initial period of about 200 fs, during which the population dynamics with and without the RWA are very similar, the corresponding curves become out of phase is a clear indication of the non-dynamical energy shift $\Delta\omega_{ndyn}$ due to the counter-rotating effects.

In the lower panel of Fig. \ref{fig3}, we present the $|c_1(t)|^2$ time evolution for a QE with
$\omega_0=1.959456$ eV; the inset in the panel shows the dynamics for a QE with $\omega_0=1.97262$ eV.
The computed  $|c_1(t)|^2$ time evolution for a QE with $\omega_0=1.959456$ eV, as well as for a QE with $\omega_0=1.97262$ eV, is distinctly different than the dynamics presented in the upper panel of this figure.  In the lower main panel, we observe non-Markovian dynamics during which the population of the QE oscillates back and forth to the electromagnetic  mode continuum, while the QE initial population decays out totally within 3-4 ps. On top of these oscillations, small amplitude high-frequency oscillations are superimposed.

Similar dynamics is also observed in the inset of this panel for the QE with $\omega_0=1.97262$ eV; the only difference to the dynamics for the QE with $\omega_0=1.959456$ eV  is that now the decay is slightly faster and the superimposed oscillations on the overall decay have slightly larger amplitude. As in the case of the dynamics shown in the upper panel of this figure, such high-frequency oscillations are due to the overlap of the exciton-polariton resonances closest to the QE resonance frequency due to the combined effect of the
dynamical energy shift and the $\Delta\omega_{ndyn}\approx 0.3$ meV, valid for both $\omega_0$ shown in
this panel. However, in the cases presented in the lower panel of Fig. \ref{fig3}, such overlap is more extensive than in the case shown in the upper panel of this figure, resulting to more pronounced oscillatory features on top of the overall non-Markovian population time evolution. As in the upper panel of this figure, the non-dynamical energy shift, which is only due to the counter-rotating effects, renders the population dynamics time-evolution with and without applying the RWA out of phase after a short  initial period.

We note that when the QE is located at $z=5$ nm, due to the fact that the exciton-polariton resonances
are overlapping, any single peak fitting of the $\tilde{J}(\omega)$ at this distance to a Lorentzian profile
results in a very poor approximation; consequently, the analytical solution $|c^L_1(t)|^2$ for the corresponding population dynamics fails to describe the spontaneous emission  process even qualitatively in comparison to the corresponding exact $|c_1(t)|^2$ time evolution (not shown here).

\begin{figure}[h]
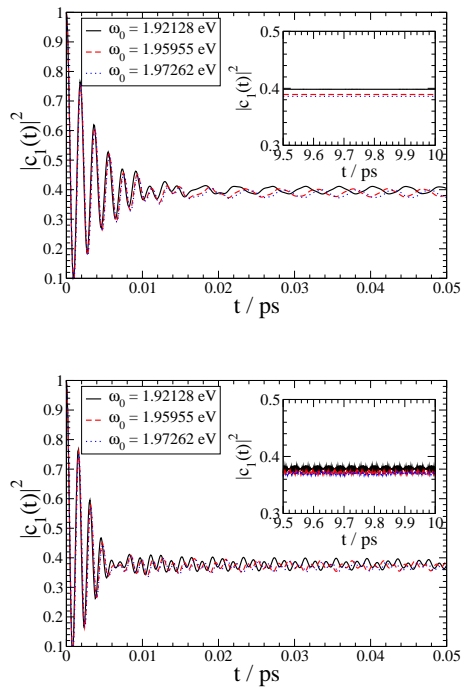

\centerline{\hbox{\epsfxsize=60mm\epsffile{fig4a.eps}}}
\vspace*{0.7cm}
\centerline{\hbox{\epsfxsize=60mm\epsffile{fig4b.eps}}}
\begin{center}
\caption{\label{fig4} (color online) Population dynamics  of a QE without the RWA (upper panel) and with the RWA (lower panel) with a $x$-oriented transition dipole moment, $\Gamma_0(\omega_0)=413.5$ $\mu$eV and resonance frequency  $\omega_0$  located at $\mathbf{\vec{r}}_{\text{QE}}=(0,0,2)$ nm next to a high-quality MoS$_2$ disk  with radius $R=30$ nm. The inset in each panel presents the dynamics as in the main panel at later times. }
\end{center}
\end{figure}

In Fig. \ref{fig4} we show the dynamics under ultrastrong coupling conditions \cite{reviewultra} between the QE and the electromagnetic  mode continuum modified by the nanodisk. For this purpose, we investigate a QE with free-space decay width
$\Gamma_0(\omega_0)=413.5$ $\mu$eV, such as a J-aggregate used in recent experiments \cite{liu17}; the QE
is located at $z=2$ nm from a high-quality MoS$_2$ nanodisk of 30 nm radius.  The population dynamics is calculated for three QE resonance frequencies $\omega_0$; the same frequencies as in Fig. \ref{fig3}.  We present the dynamics including counter-rotating effects (upper panel) as well as invoking the RWA (lower panel).
The $\Delta\omega_{ndyn}$ values corresponding to the results shown in the upper panel are in all cases about
40 meV at this distance, qualitatively indicating ultrastrong light-matter coupling conditions. In the main part of each panel of Fig. \ref{fig4}, we observe in all $\omega_0$ cases shown, a very fast oscillatory decay of the QE population within the first 10 fs to about 0.4 of its initial value; after this early stage, we find that this population remains practically constant, indicating population trapping in the QE. The period of oscillations in the early stage of dynamics is about 2 fs and is practically the same as $2\pi/\omega_{0}$, showing clearly that the dynamics is in the ultrastrong coupling regime \cite{reviewultra}. Also, the population trapping occurring here is attributed to the creation of a bound state of the QE - localized exciton-polariton modes due to the ultrastrong coupling, as analyzed in Ref. \cite{Yang17a}.   The inset in both panels of this  figure shows the population dynamics shown in the main part of the corresponding panel at much later times, during 9.5-10 ps, demonstrating that the QE population is constant over long times. We note that invoking the RWA leads to very small-amplitude  high-frequency oscillations on the value of the trapped population, which persist even at long times, in contrast to the dynamics including counter-rotating effects.

\begin{figure}[h]
\centerline{\hbox{\epsfxsize=70mm\epsffile{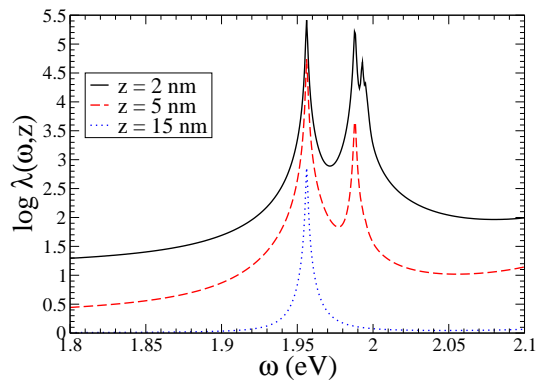}}}
\begin{center}
\caption{\label{fig5} (color online) Purcell factor $\lambda(\omega,z)$ of a QE with $x$-oriented transition dipole moment located at $\mathbf{\vec{r}}_{\text{QE}}=(0,0,z)$ nm next to a high quality MoS$_2$ nanodisk of radius $R=7.5$ nm.}
\end{center}
\end{figure}

We now focus on studying the influence of the nanodisk size on the non-Markovian spontaneous emission dynamics of the QE. In Fig. \ref{fig5}  we explicitly show the Purcell factors $\lambda(\omega,z)$ along the $x$ direction  for a QE located at $z = 2, 5, 15$ nm from a high-quality MoS$_2$ nanodisk of radius  $R=7.5$ nm. As in case of the $R=30$ nm radius nanodisk, a significant enhancement of the Purcell factor is observed for several orders of magnitude. The strong and sharp peaks represent the exciton-polariton resonances. However, at each $z$, the number of peaks is smaller than in the corresponding case for the larger disk; in addition, the exciton-polariton resonances are distinctly less overlapping than in case of the $R=30$ nm radius nanodisk.

\begin{figure}[h]
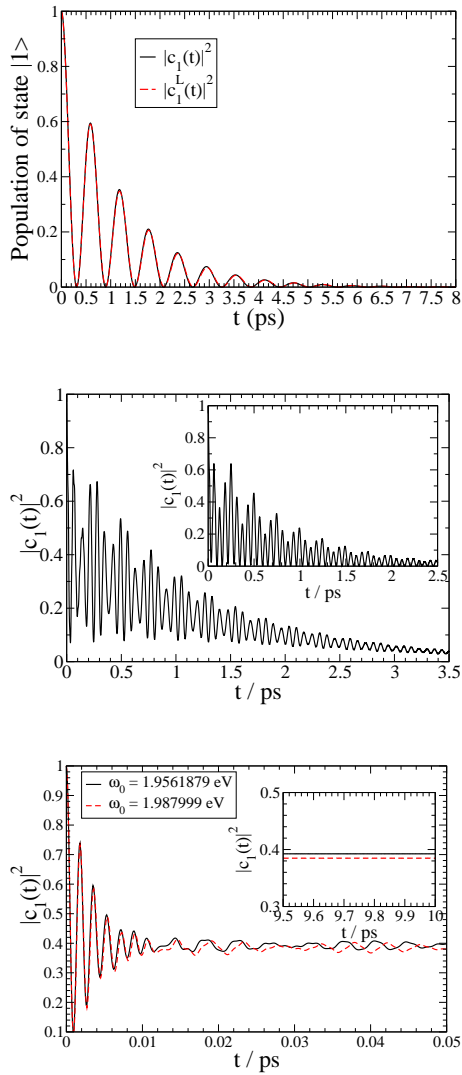

\centerline{\hbox{\epsfxsize=60mm\epsffile{fig6a.eps}}}
\vspace*{0.7cm}
\centerline{\hbox{\epsfxsize=60mm\epsffile{fig6b.eps}}}
\vspace*{0.7cm}
\centerline{\hbox{\epsfxsize=60mm\epsffile{fig6c.eps}}}
\begin{center}
\caption{\label{fig6} (color online) Population dynamics without the RWA of a QE with a $x$-oriented transition dipole moment,  next to a high quality MoS$_2$ disk  with radius $R=7.5$ nm. Top panel:  The QE with resonance frequency $\omega_0=1.956165$ eV and $\Gamma_0(\omega_0)=59$ $\mu$eV is located at
$\mathbf{\vec{r}}_{\text{QE}}=(0,0,15)$ nm.  The $|c^L_1(t)|^2$ (red dotted curve) and the $|c_1(t)|^2$ (black solid curve) are shown. The $|c^L_1(t)|^2$ curve is obtained by assuming $\omega_c=\omega_0$.
Middle panel: The QE with $\Gamma_0(\omega_0)=59$ $\mu$eV and resonance frequency
$\omega_0=1.9561879$ eV (main panel) and $\omega_0=1.987999$ eV (inset)  located at $\mathbf{\vec{r}}_{\text{QE}}=(0,0,5)$ nm. Bottom panel: The QE with resonance frequency $\omega_0$ and $\Gamma_0(\omega_0)=413.5$ $\mu$eV is located at $\mathbf{\vec{r}}_{\text{QE}}=(0,0,2)$ nm. }
\end{center}
\end{figure}

In Fig. \ref{fig6} we present the population dynamics without invoking the RWA of a QE located at $z=15$ nm
(top panel), $z=5$ nm (middle panel) and $z=2$ nm (bottom panel)  next to a high-quality
MoS$_2$ nanodisk with $R=7.5$ nm; for the first two cases, we have $\Gamma_0(\omega_0)=59~\mu$eV, and
for the latter, we have $\Gamma_0(\omega_0)=413.5~\mu$eV. In all three cases, we observe similar population time evolution to the corresponding cases of the QE next to the larger disk. More specifically,  in the upper panel of this figure, we investigate a QE with $\omega_0=1.956165$ eV, which corresponds to the single exciton-polariton resonance of the nanodisk at this distance. We observe distinct non-Markovian features in the population dynamics (black solid curve). The population of the upper QE level  has Rabi oscillations with gradually decreasing amplitude before it totally decays out of the QE within 3 to 4 ps.  In the same panel of Fig. \ref{fig6}, we also present the upper level population dynamics of the QE $|c^L_1(t)|^2$ (red dotted curve) after fitting the $\tilde{J}(\omega)$ for a QE at $z=15$ nm using the  Lorentzian profile $J^L(\omega)$ with $\omega_c=\omega_0$ \cite{betas}. The population dynamics $|c^L_1(t)|^2$  is almost indistinguishable from the dynamics  computed by using $\tilde{J}(\omega)$ as obtained from the electromagnetic Green's tensor calculations.

In the middle panel of Fig. \ref{fig6} we show the population dynamics including counter-rotating effects for
a QE with resonance frequencies $\omega_0=1.9561879$ eV (main panel) and $\omega_0=1.987999$ eV (inset) located at $z=5$ nm from a high-quality MoS$_2$ nanodisk with $R=7.5$ nm. These $\omega_0$ correspond to the two exciton-polariton resonances in the nanodisk. The population dynamics time evolution in both cases has similar qualitative features as the corresponding dynamics  of a QE next to the larger nanodisk at the same distance presented in Fig. \ref{fig3}. The dynamics at $\omega_0=1.9561879$ eV is clearly non-Markovian; the QE population oscillates rapidly back and forth in the modified by the nanodisk mode continuum, while in total it decays out within 4 ps. We note that the population oscillations between the QE and the electromagnetic mode continuum are only in part, indicating that the effective QE resonance frequency is rather off-resonant to the nanodisk exciton-polariton at  $\omega=1.9561879$ eV. As discussed  above in case of the larger disk, there is a combined shift induced on the $\omega_0$ of the QE, which is dynamical as well as non-dynamical in origin; the $\Delta\omega_{ndyn}$ in this case is 0.236 meV. The dynamics in case of $\omega_0=1.987999$ eV (inset) are similar to the dynamics in case of $\omega_0=1.9561879$ eV; the main difference is that the population oscillation back and forth in the electromagnetic  mode continuum is complete during the total QE population decay, which now takes place faster, within about 3 ps. The $\Delta\omega_{ndyn}$ in this case is 0.229 meV.

In the bottom panel of Fig. \ref{fig6} the population dynamics of a QE with $\omega_0=1.9561879$ eV (black solid curve) and $\omega_0=1.987999$ eV (red dashed curve) located at $z=2$ nm from a high-quality
MoS$_2$ nanodisk with $R=7.5$ nm is shown. At this distance between the QE and the nanodisk, the coupling between them is ultrastrong; the corresponding  $\Delta\omega_{ndyn}$ is about 43 meV in both $\omega_0$ cases. Such coupling conditions result into trapping of the QE population, as in the corresponding case next to a larger disk presented in the upper panel of Fig. \ref{fig4}. Here, for both $\omega_0$, there is a short initial period of about 10 fs, during which the population oscillates with decreasing amplitude before it reaches a constant value of about 0.4 of the initial population, which persists even at much longer times, 9.5-10 ps, as shown in the inset of this panel. The population oscillation at early times has a period which is smaller than $2\pi/\omega_{0}$, showing that the interaction is in the ultrastrong coupling regime.
We note that the population dynamics with the RWA for the cases shown in Fig. \ref{fig6} are not shown here, since the results are qualitatively similar to  the corresponding cases of a QE next to a larger disk presented
in Fig. \ref{fig3}.

\begin{figure}[h]
\centerline{\hbox{\epsfxsize=70mm\epsffile{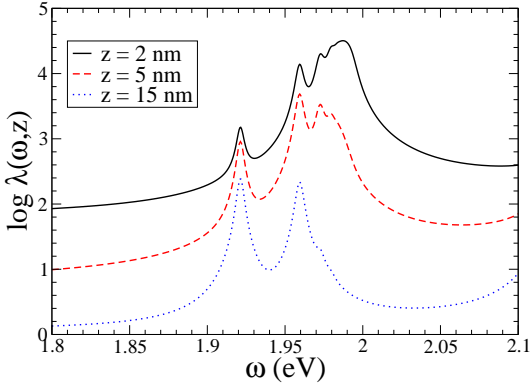}}}
\begin{center}
\caption{\label{fig7} (color online) Purcell factor $\lambda(\omega,z)$ of a QE with $x$-oriented transition dipole moment located at $\mathbf{\vec{r}}_{\text{QE}}=(0,0,z)$ nm next to a low-quality MoS$_2$ nanodisk
of radius $R=30$ nm.}
\end{center}
\end{figure}
We lastly investigate the influence of the material quality on the spontaneous emission dynamics next the  MoS$_2$ nanodisk. For this purpose, we discuss next the dynamics of a QE in proximity to a low-quality nanodisk with $R=30$ nm. In Fig. \ref{fig7}  we explicitly show the Purcell factors $\lambda(\omega,z)$ along the $x$ direction  for a QE located at $z = 2, 5, 15$ nm from a low-quality MoS$_2$ nanodisk of radius  $R=30$ nm.
As in case of the high-quality nanodisk, a significant enhancement of the Purcell factor is observed for several orders of magnitudes. The several peaks represent the exciton-polariton resonances. However, only when the QE is located at large distance $z=15$ nm from the nanodisk, all peaks are distinct, although overlapping to great extend; at closer distances only one peak at $\omega=1.92128$ eV remains distinct, while the other peaks merge into a broad resonance structure due to strong overlapping between them. Therefore, we focus on QE with
$\omega_0=1.92128$ eV in the dynamical investigation below.

\begin{figure}[h]
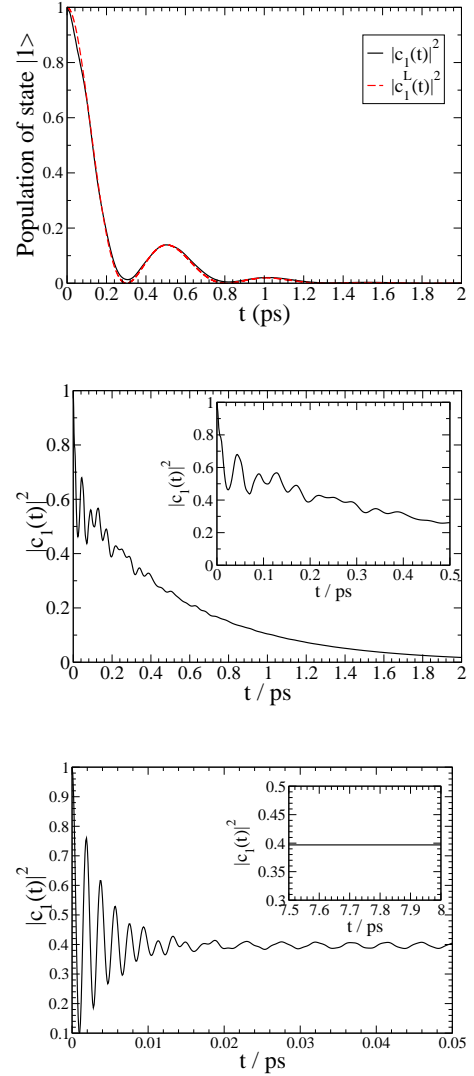

\centerline{\hbox{\epsfxsize=60mm\epsffile{fig8a.eps}}}
\vspace*{0.7cm}
\centerline{\hbox{\epsfxsize=60mm\epsffile{fig8b.eps}}}
\vspace*{0.7cm}
\centerline{\hbox{\epsfxsize=60mm\epsffile{fig8c.eps}}}
\begin{center}
\caption{\label{fig8} (color online) Population dynamics without the RWA of a QE with resonance frequency $\omega_0=1.92128$ eV and $x$-oriented transition dipole moment next to a low quality MoS$_2$ disk  with radius $R=30$ nm. In the top panel we present both $|c_{1}(t)|^2$, without applying the RWA, (black solid curve) and $|c^L_1(t)|^2$ (red dotted curve), while in the other two panels we present only $|c_{1}(t)|^2$. Top panel:  The QE with $\Gamma_0(\omega_0)=59$ $\mu$eV is located at $\mathbf{\vec{r}}_{\text{QE}}=(0,0,15)$ nm. Middle panel: The QE  with $\Gamma_0(\omega_0)=59$ $\mu$eV is located at
$\mathbf{\vec{r}}_{\text{QE}}=(0,0,5)$ nm. The inset presents the dynamics as in the main panel at early times.      Bottom panel: The QE  with $\Gamma_0(\omega_0)=413.5$ $\mu$eV is located at $\mathbf{\vec{r}}_{\text{QE}}=(0,0,2)$ nm. The inset presents the dynamics as in the main panel at later times. }
\end{center}
\end{figure}

In Fig. \ref{fig8} we present the population dynamics without applying the RWA of a QE with
$\omega_0=1.92128$ eV located at $z=15$ nm (top panel), $z=5$ nm (middle panel) and $z=2$ nm
(bottom panel)  next to a low-quality MoS$_2$ nanodisk with $R=30$ nm; for the first two cases, we
have $\Gamma_0(\omega_0)=59~\mu$eV, and for the latter, we have $\Gamma_0(\omega_0)=413.5~\mu$eV. More specifically,  in the upper panel of this figure, the QE is located at large distance, $z=15$ nm, from the nanodisk, resulting into non-Markovian decay dynamics, which however is less pronounced than in the corresponding case of a QE next to a high-quality nanodisk with $R=30$ nm shown in Fig. \ref{fig2}, and the population decays totally within 1.5 ps. In this case, too, the analytical solution $|c^L_1(t)|^2$ for the  population dynamics, after the applying the fitting procedure \cite{betas} in this case here, gives a good approximation in comparison to the corresponding exact $|c_1(t)|^2$ time evolution.
The population dynamics including counter-rotating effects of a QE at $z=5$ nm next to the low-quality nanodisk of 30 nm radius is shown in the middle panel of this figure. The inset of this panel presents the same dynamics as the main panel at early times. We observe that the dynamics has non-Markovian features, which are much stronger in the first 200 fs. After that time, these features weaken, resulting to a gradual decay, which is completed within 2 ps. This fact is due to the wide exciton-polariton resonance of the nanodisk at $\omega=1.92128$ eV, as well as its strong overlap with the broad resonance structure at higher energies. The $\Delta\omega_{ndyn}$ in this case is 0.3 meV.

In the bottom panel of Fig. \ref{fig8}, the population dynamics without invoking the RWA of a QE located at
$z=2$ nm from a low-quality MoS$_2$ nanodisk with $R=30$ nm is presented. Interestingly, despite the low-quality of the MoS$_2$ nanodisk discussed here, we observe that the population oscillates rapidly  with decreasing amplitude between the QE and the electromagnetic  mode continuum during the first 10 fs with a period of oscillation about $2\pi/\omega_{0}$, while after that period the QE population is partly trapped as indicated by the practically constant value of about 0.4 of the initial population. In the inset of the same panel the population time evolution is presented at much longer times, 7.5 to 8 ps, indicating the persistent character of the population trapping under the present coupling conditions. We note that the  $\Delta\omega_{ndyn}$ in this case is 39 meV.

\section{Summary}
In summary, we investigate the spontaneous emission  dynamics of a two-level QE in proximity to a MoS$_2$ nanodisk. The decay dynamics has strong non-Markovian character at close distance between the QE and the nanodisk. In this case significant exciton-polariton resonance overlapping is affecting the population dynamics, in combination with the intensity of the counter-rotating effects, leading to either complex decaying oscillations or ultrastrong coupling and population trapping. As the distance increases, the population evolution gives clear decaying Rabi oscillations, as the effect of the counter-rotating effects becomes marginal in addition to the vanishing of exciton-polariton resonance overlapping.  Similar results are also obtained for nanodisks with different radius, as well as lower material quality nanodisks. We  believe that our results are useful for the development of novel devices on the nanoscale and future quantum technologies.

\section*{Acknowledgments}
Co-financed by Greece and the European Union-European Regional Development Fund via the General Secretariat for Research and Technology (GSRT) (project POLISIMULATOR).
N.I. acknowledges the support of his Ph.D. by the General Secretariat for Research and Technology (GSRT) and the Hellenic Foundation for Research and Innovation (HFRI) via a doctoral scholarship (Grant No. 2649).

\end{document}